\documentclass[journal,10pt,twocolumn]{IEEEtran}
\usepackage[pdftex]{graphicx}
\DeclareGraphicsExtensions{.pdf,.jpeg,.png}
\DeclareGraphicsExtensions{.eps}
\usepackage{graphicx}
\usepackage{epstopdf}
\usepackage{amsmath}

\usepackage{amssymb}
\usepackage{algorithmic}
\usepackage{cite}
\usepackage{subfig}
\usepackage{caption}
\usepackage{lipsum}
\usepackage{amsthm}
\newtheorem{theorem}{Theorem}

\newtheorem{lemma}{Lemma}
\renewcommand{\qedsymbol}{$\blacksquare$}
\begin{document}
\title{Analytical Model for Outdoor Millimeter Wave Channels using Geometry-Based Stochastic Approach}
\author{Nor~Aishah~Muhammad,~\IEEEmembership{Student Member,~IEEE,}
        Peng~Wang,~\IEEEmembership{Member,~IEEE,}
				Yonghui~Li,~\IEEEmembership{Senior~Member,~IEEE,}
        and~Branka~Vucetic,~\IEEEmembership{Fellow,~IEEE}
\thanks{Copyright (c) 2015 IEEE. Personal use of this material is permitted. However, permission to use this material for any other purposes must be obtained from the IEEE by sending a request to pubs-permissions@ieee.org.}
\thanks{N. A. Muhammad is with the School of Electrical and Information Engineering, The University of Sydney, NSW 2006, Australia, and also with the Wireless Communication Centre, Faculty of Electrical Engineering, UTM Johor Bahru, Malaysia (e-mail: nor.muhammad@sydney.edu.au).}
\thanks{P. Wang (Corresponding author) is with Huawei Technologies Sweden AB, Kista, Sweden. His work in this paper was partially done when he was with the School of Electrical and Information Engineering, The University of Sydney, NSW 2006, Australia. (e-mail: wp\_ady@hotmail.com).}
\thanks{Y. Li and B. Vucetic are with the School of Electrical and Information Engineering, The University of Sydney, NSW 2006, Australia (e-mail: yonghui.li@sydney.edu.au; branka.vucetic@sydney.edu.au).}}

\markboth{IEEE TRANSACTIONS ON VEHICULAR TECHNOLOGY, ACCEPTED FOR PUBLICATION}%
{Shell \MakeLowercase{\textit{et al.}}: Bare Demo of IEEEtran.cls for Journals}

\maketitle
\begin{abstract}
\boldmath
The severe bandwidth shortage in conventional microwave bands has spurred the exploration of the millimeter wave (MMW) spectrum for the next revolution in wireless communications. However, there is still lack of proper channel modeling for the MMW wireless propagation, especially in the case of outdoor environments. In this paper, we develop a geometry-based stochastic channel model to statistically characterize the effect of all the first-order reflection paths between the transmitter and receiver. These first-order reflections are generated by the single-bounce of signals reflected from the walls of randomly distributed buildings. Based on this geometric model, a closed-form expression for the power delay profile (PDP) contributed by all the first-order reflection paths is obtained and then used to evaluate their impact on the MMW outdoor propagation characteristics. Numerical results are provided to validate the accuracy of the proposed model under various channel parameter settings. The findings in this paper provide a promising step towards more complex and practical MMW propagation channel modeling. 
\end{abstract}
\section{Introduction}
\IEEEPARstart{T}{he} tremendous growth of wireless services on hand-held devices, such as high-definition video streaming, online gaming and cloud computing, has led to an explosive demand for mobile data~\cite{Cisco2016}. Nowadays, wireless service providers are struggling to deliver high-quality wireless networks by utilizing sophisticated modulation schemes and signal processing technologies in a quest for a higher data rate. However, achieving a transmission rate up to the order of Gigabits/second (Gbps) or even higher is quite challenging due to the limited bandwidth available at microwave frequencies. Therefore, more bandwidth is required to alleviate the global bandwidth shortage. Recently, researchers have moved their attention to the underutilized millimeter wave (MMW) signals~\cite{Geng2009,Rappaport2013,Haneda2014,Akdeniz2014,Metis2014,Zhang2010,Peng2014} ranging from 30 GHz to 300 GHz, as these bands provide much wider spectrum resources than their microwave counterparts.
Although MMW bands potentially offer numerous significant performance improvements in wireless networks, including extremely high speed and low latency services, they still face many technical challenges related to the unique propagation characteristics of MMW bands. For example, MMW signals encounter severe path losses, high atmospheric and rain attenuation, making the deployment of MMW systems very challenging, especially for outdoor communications\cite{Peng2015,Rappaport2013, Andrew5G2014}. In addition, MMW signals cannot penetrate most solid materials such as buildings, resulting in the isolation of outdoor base stations from indoor users~\cite{Zhouyue2011}. Furthermore, sharp shadow zones occur due to the significantly larger sizes of buildings relative to the signal wavelength at MMW frequencies, which lead to insignificant diffraction mechanisms that can be neglected in the MMW system analysis~\cite{Haneda2014,HaoXu2002}. Thus, MMW channels are expected to exhibit a sparse multipath nature, instead of the rich-scattering one demonstrated in conventional microwave channels\cite{Zhouyue2011}. As a result, it is not possible to directly use microwave propagation models for MMW systems. An in-depth understanding of the MMW propagation characteristics is essential for the design and analysis of future MMW wireless networks.
\subsection{Related Work}
There have been some research efforts on modelling the MMW propagation channel~\cite{Priebe2013_2,Geng2009,Rappaport2013,Haneda2014,Akdeniz2014,Metis2014,HaoXu2002,Zhang2010}. Extensive propagation measurements have been conducted in~\cite{Geng2009,Rappaport2013,Haneda2014}, which reveal the unique characteristics of wireless propagation at MMW bands. Motivated by these characteristics, several attempts have been done on modeling the MMW channels, including deterministic and stochastic models. Deterministic models are typically based on ray tracing simulations~\cite{Zhang2010, Larew2013}, which aim to produce accurate results via a detailed description of propagation environments. So they are costly and time-consuming especially when a large investigation area is considered. In addition, such results would be valid only for the particular propagation setting and may not be applicable to general propagation environments.

On the other hand, the stochastic approach that characterizes the channel behaviour using the probability distribution functions of the channel parameters is becoming a popular way to develop general yet sufficiently accurate channel models. Stochastic channel models can be further classified into two approaches, which are non-geometrical\cite{Geng2009} and geometry-based~\cite{Akdeniz2014,Metis2014}. In~\cite{Geng2009}, the authors proposed a non-geometrical channel model to statistically characterize the channel parameters, such as the number of scatterers, delay spread, path loss and shadowing, without any geometric assumptions. In contrast, the geometry-based channel models in~\cite{Akdeniz2014,Metis2014} were developed based on the predefined distributions of the channel parameters and distribution of effective scatterers with their geometric information such as angles of departure and arrival, and delay. All the parameters in~\cite{Geng2009,Akdeniz2014,Metis2014} were obtained from an extensive set of channel measurements. However, parameterization of these models are currently lacking because of the limited MMW channel measurement data.

The geometry-based stochastic approach has also been adopted in analytical channel models~\cite{Marano1999,Marano2005, Bai2014}. Specifically, the authors in~\cite{Marano1999} modeled all the buildings in an urban area as random lattices with a constant occupancy probability, which refers to the probability that a lattice area is occupied by a building. A closed-form expression for the propagation depth, i.e., the probability that a ray undergoes the successive reflection steps at a certain level of lattices, has been proposed based on the assumption that the ray enters the lattice area at a prescribed incident angle. In~\cite{Marano2005}, the lattice environment in~\cite{Marano1999} was extended by removing the restriction of the incident angle. Such a refinement was achieved by assuming that the transmitter was placed inside the lattice. With such a system model, the authors considered the possibility of the successive reflections at all propagation angles. However, the existing models in~\cite{Marano1999,Marano2005} have been mostly designed for the rich scattering environment, which will not be valid for the sparse MMW channels.    

More recently, the authors in~\cite{Bai2014} proposed a stochastic geometry approach for modeling the outdoor MMW propagation environment, where the locations of the buildings are assumed to follow a Poisson Point Process (PPP) and the shapes of the buildings are modeled as random rectangles. A distance-dependent probability of a line-of-sight (LoS) link has been proposed and used in the network-level performance analysis. In~\cite{Singh2015}, the authors proposed a simplified LoS probability, which is characterized by the average fraction of LoS area around the point under consideration. The works in~\cite{Bai2014, Singh2015} mainly focused on the blockage effects of buildings without considering their contributions on multipath components. However, as verified by the channel measurements in~\cite{Rajagopal2012}, the reflection paths generated by buildings can also provide a non-negligible multipath propagation, which highlights the importance of considering the reflection paths in MMW channel modeling. So far, there has been only a few discussions regarding the reflection paths for MMW channels, including~\cite{Zhang2010, Clara2014, Samimi2014, Geng2009}. The common limitation of these works is that the randomness of the locations, sizes and orientations of buildings were not explicitly incorporated, making these models not applicable to general environments. To the best knowledge of the authors, an analytical characterization of multipath components incorporating the MMW features and the randomness of the propagation environment has not yet been developed.

\subsection{Contributions}
In this paper, we develop a stochastic MMW channel model by considering all the first-order reflection components\footnote{Since higher-order reflections in MMW systems suffer from more severe path losses and attenuation, one would expect that the power contributed by these reflections is insignificant and can be neglected. Hence it is reasonable to mainly focus on the first-order reflection paths in the analysis as these paths provide the strongest received power among all the reflection components~\cite{Jarvelainen2014}. A similar treatment has also been adopted in~\cite{Oestges2000,Raisanen2013,Jarvelainen2014}.} of MMW signals in an urban area. The urban area is modeled based on the stochastic geometry approach similar to the one in~\cite{Bai2014}. On the basis of the proposed propagation environment model, we derive a closed-form expression for the power delay profile (PDP) contributed by all the first-order reflection paths, which gives the intensity of the energy at the receiver relative to the propagation delay. Then, based on the PDP, we analyze the average path loss and the number of the first-order reflection paths under various channel parameter settings. Our analytical results show that the richness of the first-order reflection paths relies on the propagation environments such as the sizes and density of buildings. The contribution of the first-order reflection paths is comparable to the LoS path, especially in dense building areas. Meanwhile, in sparse building environments, the contribution of the first-order reflection paths generated by small buildings is higher than those produced by large buildings. Our findings in this paper provide useful insights into the MMW propagation channels which can be utilized in the system performance analysis.

This paper is organized as follows. The system model is introduced in Section II. In Section III, we present the analytical derivation of the closed-form expression for the PDP contributed by all the first-order reflection paths. Section IV provides extensive discussions of our observations and the comparison between the derived formula and the numerical simulations. Final comments and conclusion are drawn in Section V.
\section{System Model}\label{sec:SModel}
\begin{figure}[!t]
\centering
\vspace{-1em}
\includegraphics[scale=0.5]{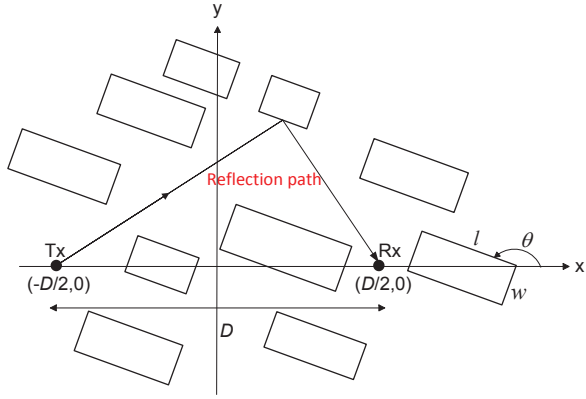}
\vspace{-1em}
\caption{System model for outdoor communications.}
\label{fig:1}
\end{figure}

Consider an MMW communication link with a separation distance $D$ between the transmitter Tx and receiver Rx, both of which are equipped with a singe omnidirectional antenna with unit gain\footnote{Although large antenna arrays are typically equipped in practical MMW cellular systems, we only consider a single omnidirectional antenna system here to provide an antenna-independent channel model, which is desirable for the purpose of system analysis. A similar treatment has also been adopted in~\cite{Bai2014,Samimi2014,Akdeniz2014}. One would obtain the PDP with directional antenna by computing the convolution of the PDP with omni-directional antenna and the antenna geometries.}. As illustrated in Fig.~\ref{fig:1}, the transmitter and receiver are located at $(-D/2,0)$ and $(D/2,0)$, respectively, and the communication link is surrounded by buildings that are randomly distributed in the communication area. For simplicity, we ignore the heights of all buildings and so a two-dimensional coordinate system as used in Fig.~\ref{fig:1} is sufficient to describe  the system. In addition, the following assumptions are made for all the buildings throughout this paper.
\begin{enumerate}
\item\label{A1} Each building is of a rectangular shape specified by its center location $C$, length $l$, width $w$ and orientation $\theta$, where $\theta$ is defined as the anti-clockwise angle between the $x$-axis and the $l$-side of the building as shown in Fig.~\ref{fig:1}. 
\item\label{A2} All the building centers $\left\{C\right\}$ form a homogeneous PPP with density $\lambda$.
\item\label{A3} The lengths and widths of all buildings follow, respectively, independent and identical distributions $f_L(l)$ and $f_W(w)$.
\item\label{A4} In each channel realization, the orientations of all buildings are the same, following a uniform distribution\footnote{This assumption is reasonable for modern cities where the buildings are normally aligned at the same angle, but may not be accurate in old cities. A further investigation for the area with different building orientations is left as a future work.} over $(0, \pi]$.
\item\label{A5} The surfaces of all buildings are sufficiently smooth such that all the non-LoS (NLoS) paths follow the specular reflection law and the diffraction paths are negligible as this propagation mechanism contributes to insignificant signal strength at MMW bands~\cite{Haneda2014,HaoXu2002}.
\end{enumerate}
In this paper, our main focus is on the characterization of the channel PDP contributed by all the first-order reflection paths.
\section{Power Delay Profile Characterization}\label{sec:PDP}
\subsection{Derivation of PDP}\label{sec:PDP_sub}\label{sec:fSR}
In the above-considered system, the signal emitted from the transmitter may arrive at the receiver via multiple propagation paths. Besides a direct LoS path, any building in the communication area may potentially generate a reflection path, depending on its location and orientation. Meanwhile, each potential reflection path may also be blocked by some other buildings. Denote by $L_r$ the propagation distance of a first-order reflection path, which corresponds to a propagation delay of $\tau=L_r/c$ with $c$ being the speed of light. The path loss along a reflection path with propagation delay $\tau$ is obtained by the Friss free space equation as~\cite{Parson2000}
\begin{equation}\label{eq:2}
\rho(\tau)=\left(\frac{\lambda_f}{4\pi L_r}\right)^2\cdot\frac{1}{\sigma}=\frac{1}{\left(4 \pi f\tau\right)^2\sigma}
\end{equation}
where $\lambda_f=c/f$ is the signal wavelength corresponding to the operating frequency $f$, and $\sigma$ denotes the mean value of the reflection loss. The reflection loss needs to be considered here because in reality, when a signal is reflected, there must be some power loss due to the absorption by the medium. Note that in principle the value of the reflection loss depends on the incident angle, the properties of the building material, (i.e., the dielectric constant and the conductivity), the polarization of the incident wave, and the carrier frequency. However, taking all these effects into consideration will significantly complicate the analysis. For simplicity, in this paper we use a constant mean value of the reflection loss $\sigma$ and leave the detailed treatment of the reflection loss for future work. The mean value of $\sigma$ is  determined by exploiting the empirical model as in~\cite{Kyro2013, Langen1994}. A similar treatment has also been adopted in \cite{Hansen2003, Raisanen2013, Clara2014}.

Denote by $\mathbb{E}(N_{RF}|\tau_1\leq\tau\leq\tau_2)$ the average number of the first-order reflection paths whose delays are between $\tau_1$ and $\tau_2$. The density of the first-order reflection paths with time delay $\tau$ can be obtained by 
\begin{equation}
f_{SR}(\tau)=\lim_{|\tau_2-\tau_1| \to 0}\frac{\mathbb{E}(N_{RF}|\tau_1\leq\tau\leq\tau_2)}{|\tau_2-\tau_1|}.
\end{equation}
Then the average path loss of all the first-order reflection paths with a specific common propagation delay $\tau$ can be modeled as   
\begin{equation}\label{eq:1}
P(\tau) = \rho(\tau)\cdot f_{SR}(\tau).
\end{equation}
In this paper, we will derive an analytical expression for $f_{SR}(\tau)$ by considering the random locations, sizes, and orientations of buildings, as detailed below.

In real communication environments, a first-order reflection path with delay $\tau$ may be generated by a building with an arbitrary orientation $\theta$. Thus, we can express the function $f_{SR}(\tau)$ as 
\begin{equation}\label{eq:3}
f_{SR}(\tau) = \int_0^{\pi} f_{SR}(\tau|\theta)\cdot f_{\theta}(\theta) \,d\theta
\end{equation} 
where $f_{SR}(\tau|\theta)$ is the density of the first-order reflection paths with time delay $\tau$ when the building orientation is $\theta$, and  $f_{\theta}(\theta)$ is the probability density function (pdf) of orientation $\theta$. From assumption~\ref{A4}, we have $f_{\theta}(\theta)=1/\pi$.   

\begin{figure}[t]\centering
\vspace{-1em}
\includegraphics[scale=0.8]{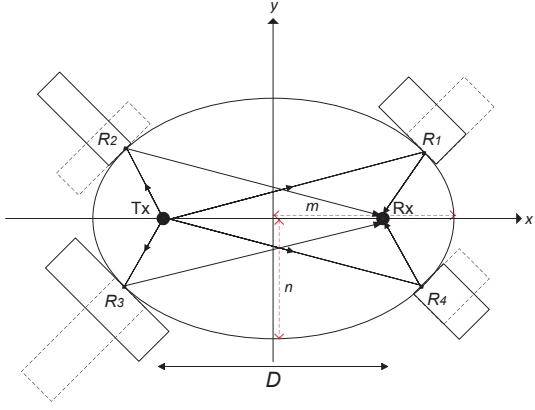}
\vspace{-1em}
\footnotesize\caption{ Examples of the reflection points on an ellipse with path length $c\tau$. All buildings in dashed and solid lines are, respectively, at the orientation of buildings $\theta\in(0,\pi/2]$ and $\theta\in(\pi/2,\pi]$.}
\vspace{-1.5em}
\label{fig:ellipse4Q}
\end{figure}

Given the positions of the transmitter and receiver, a first-order reflection path can be completely determined by its reflection point. Since all the first-order reflection components with the same delay $\tau$ have the same path length $L_r$, their reflection points can be elegantly characterized by using an ellipse model as depicted in Fig.~\ref{fig:ellipse4Q}. The foci of this ellipse are chosen to be, respectively, the transmitter Tx and receiver Rx located at $(-D/2, 0)$ and $(D/2, 0)$. Then the reflection points corresponding to the first-order reflection paths with the same delay $\tau$ all lie on this ellipse with proper major radius $m$ and minor radius $n$, as illustrated in Fig.~\ref{fig:ellipse4Q}. Based on this ellipse model, we can readily conclude that, for a given value of building orientation $\theta$, there are four possible locations of reflection points on the ellipse, which are denoted by $R_1, R_2, R_3$ and $R_4$ in Fig.~\ref{fig:ellipse4Q}, respectively. Consequently, we have four exclusive events of the first-order reflection paths with given $\tau$ and $\theta$. Thus the function $f_{SR}(\tau|\theta)$ in (\ref{eq:3}) can be rewritten as
\begin{equation}\label{eq:3a}
f_{SR}(\tau|\theta)=\sum_{i=1}^{4}f_{SR_i}(\tau|\theta)
\end{equation} 
where $f_{SR_i}(\tau|\theta)$ is the counterpart of $f_{SR}(\tau|\theta)$ contributed by the event when the reflection point is located at $R_i$. Note that since the derivations of these four terms in (\ref{eq:3a}) are very similar to each other, our subsequent discussion will be mainly focused on the first term, $f_{SR_1} (\tau|\theta)$. In addition, we assume $\theta \in (\pi/2, \pi]$ in what follows, as the related derivations, though the same as those for the case of $\theta \in (0, \pi/2]$, involve slightly different notations.  
\begin{figure*}[ht!]
\centering
  \vspace{-2em}
	\hspace{-5em}
  \subfloat[]{\includegraphics[width=7.7cm,height=6cm]{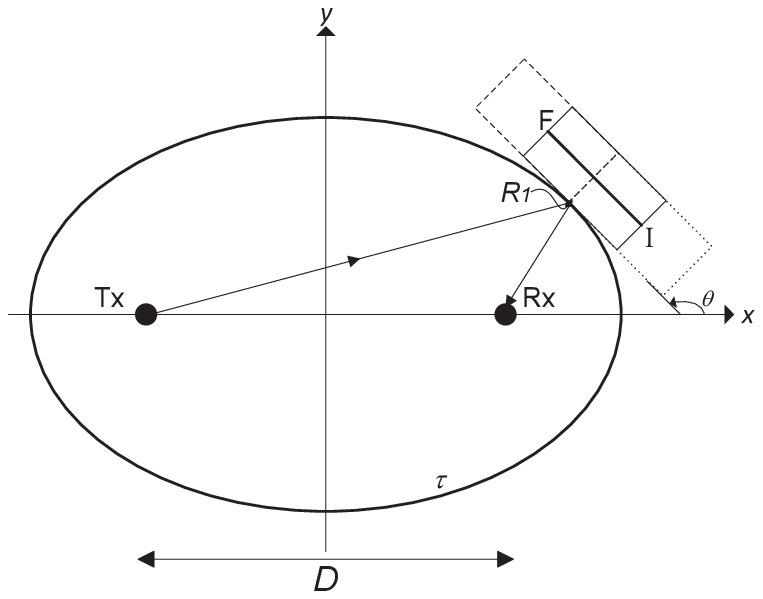}} 
	\hspace{-2.5em}
	\quad
	\hspace{-2.5em}
  \subfloat[]{\includegraphics[width=7.7cm,height=6cm]{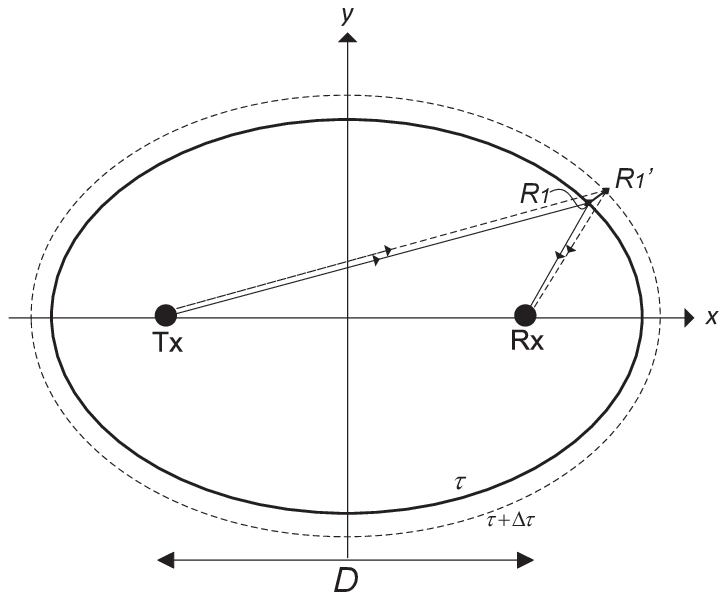}}
	\hspace{-2.5em}
	\quad
	\hspace{-2.5em}
	\subfloat[]{\includegraphics[width=6.2cm,height=5.5cm]{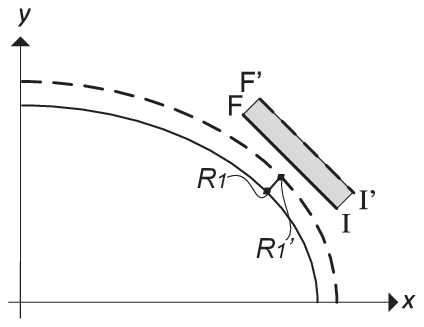}}
	\hspace{-7em}
\caption{Ellipse model. (a) The possible locations of buildings that potentially generate a reflection path with $R_1$ being its reflection point. (b) Reflection paths with path lengths $c\tau$ and $c(\tau+\Delta\tau)$.(c) Geometry of the parallelogram $FF'I'I$.}
	\label{fig:3abc}
\end{figure*}

Intuitively, a first-order reflection path that contributes to $f_{SR_1}(\tau|\theta)$ can be guaranteed if and only if the following two independent sub-events hold simultaneously: (a) there is a building with a proper location to generate such a first-order reflection path with its reflection point located at $R_1$, and (b) this reflection path is not blocked by any other buildings. Mathematically, we can further decompose the function $f_{SR1}(\tau|\theta)$ as 
\begin{equation}\label{eq:4}
f_{SR_1}(\tau|\theta)=f_{RF_1}(\tau|\theta)\cdot f_{NB_1}(\tau|\theta)
\end{equation} 
where $f_{RF_1}(\tau|\theta)$ is the density of sub-event (a) and $f_{NB_1}(\tau|\theta)$ refers to the probability of sub-event (b). The detailed derivations of them are presented in, respectively, subsections B and C below. 
\subsection{Derivation of $f_{RF_1}(\tau|\theta)$}\label{sec:REFLECTION}
To facilitate the derivation of $f_{RF_1}(\tau|\theta)$ in (\ref{eq:4}), we exploit the ellipse model as in Fig.~\ref{fig:3abc}(a) to describe the coordinates of the reflection point $R_1$, denoted by $(x, y)$. Mathematically we have 
\begin{equation}\label{derive_RF1}
\frac{x^2}{m^2}+\frac{y^2}{n^2}=1
\end{equation}  
where $m$ and $n$ are, respectively, the major radius and minor radius of the considered ellipse. Given the coordinates of the two foci, i.e., $(-D/2,0)$ and $(D/2,0)$, and the fact that the sum of the distances between any point on the ellipse and its two foci is the same and equals to $L_r=c\tau$, we have
\begin{equation}\label{m}
m=\frac{c \tau}{2},
\end{equation}
and
\begin{equation}\label{n}
n=\frac{1}{2}\sqrt{c^2\tau^2-D^2}.
\end{equation}
Recalling the assumption that $\theta \in (\pi/2, \pi]$, we can see that the tangent line of the ellipse across point $R_1$ is coincident with the $l$-side wall of the building that generates this first-order reflection path. Denote by $\varphi$ the angle between the $x$-axis and this tangent line and recall that $\theta$ represents the orientation of all buildings. From Fig.~\ref{fig:3abc}(a), we can express the slope of this tangent line by implicitly differentiating (\ref{derive_RF1}) with respect to $x$ as
\begin{equation}\label{derive_RF2}
\tan \varphi=\tan \theta = \frac{dy}{dx}=\frac{-n^2x}{m^2y},
\end{equation}
or equivalently
\begin{equation}\label{derive_RF3}
y =\frac{-n^2x}{m^2\tan \theta}.
\end{equation} 
Combining (\ref{derive_RF1}), (\ref{m}), (\ref{n}) and (\ref{derive_RF3}), we can write the coordinates of point $R_1$ as
\begin{equation}\label{derive_RF4}
x =\frac{-c^2\tau^2\tan \theta}{2\sqrt{c^2\tau^2\sec^2 \theta-D^2}},
\end{equation} 
\begin{equation}\label{derive_RF5}
y =\frac{c^2\tau^2-D^2}{2\sqrt{c^2\tau^2\sec^2 \theta-D^2}}.
\end{equation}

Now, let us assume that the above-considered reflection path is generated by a building specified by the quadruple $(C,l,w,\theta)$. As illustrated in Fig.~\ref{fig:3abc}(a), the reflection point $R_1$ may be at any point on the $l$-side wall of this building. As a consequence, the center of the building $C$ may also have numerous possible locations, and all these locations form a line segment $FI$, where the two end points $F$ and $I$ correspond to the critical cases when $R_1$ falls on the two corners of the building.

In practice, the probability of the sub-event that there is a building capable of generating a first-order reflection path with its reflection point located at a specific point, e.g., $R_1$, is always zero. Therefore, in (\ref{eq:4}), we have defined $f_{RF_1}(\tau|\theta)$ as the density, instead of the probability, of such a sub-event. To derive the expression of $f_{RF_1}(\tau|\theta)$, we need to introduce a neighbourhood of $R_1$, which is denoted by ${\mathcal{N}_{R_1}}$. Consider the following event $E_1$: conditioned on that all buildings have orientation $\theta$, there is at least one\footnote{In practice, buildings should not be overlapped, which is not guaranteed in the homogeneous PPP model in assumption~\ref{A2}. However, as the measure of the neighborhood ${\mathcal{N}_{R_1}}$ vanishes to zero, the overlapping probability will also converge to zero and thus the ignorance of overlapping will not cause any error in our model. A similar treatment has also been adopted in~\cite{Bai2014}.} properly located building that is capable of generating a first-order reflection path. Denote by $\mathbb{P}(E_1|\mathcal{N}_{R_1})$ the probability of such a reflection path with its reflection point falling within the neighbourhood of $R_1$,  ${\mathcal{N}_{R_1}}$. Then we can express $f_{RF_1}(\tau|\theta)$ as 
\begin{equation}\label{eq:5}
f_{RF_1}(\tau|\theta)=\displaystyle\lim_{|\mathcal{N}_{R_1}| \to 0}\frac{\mathbb{P}(E_1|\mathcal{N}_{R_1})}{|\mathcal{N}_{R_1}|}
\end{equation}
where $|\mathcal{N}_{R_1}|$ is the measure of ${\mathcal{N}_{R_1}}$. 

In this paper, we choose ${\mathcal{N}_{R_1}}$ to be the trajectory of $R_1$ when the path length varies from $c\tau$ to $c(\tau+\Delta\tau)$. Therefore, we have 
\begin{equation}\begin{align}\label{eq:5a}
|\mathcal{N}_{R_1}|&=c(\tau+\Delta\tau)-c\tau=c\Delta\tau.
\end{align}\end{equation}
It can be expected that, as the path length increases from $c\tau$ to $c(\tau + \Delta\tau)$, the reflection point $R_1$ will gradually move outwards to another position denoted by $R_1'$ in Fig.~\ref{fig:3abc}(b). Consequently, the line segment $FI$ will also gradually move outwards to another line segment denoted by $F'I'$ in Fig.~\ref{fig:3abc}(c). In other words, $\mathbb{P}(E_1|\mathcal{N}_{R_1})$ in (\ref{eq:5}) is equal to the probability of the event that the center of the building falls in the region swept by the line segment $FI$ during the movement. Note that this area may not be of a regular parallelogram shape, as the trajectory of the reflection point from $R_1$ to $R_1'$ may not be a straightforward line segment. However, when $\Delta \tau \to 0$, we can asymptotically regard the trajectory of $R_1R_1'$ as a line segment, and in turn regard the region $FF'I'I$ as a parallelogram, whose area is calculated in the following lemma.
\begin{lemma} When $\Delta \tau$ is sufficiently small, the area of the parallelogram $FF'I'I$ illustrated in Fig.~\ref{fig:3abc}(c), which is denoted by $S_{FF'I'I}$,  is given by
\begin{equation}\label{eq:5b}
S_{FF'I'I}=\frac{l c^2\tau\Delta \tau}{2\sqrt{c^2\tau^2-D^2\cos^2\theta}}.
\end{equation} 
\end{lemma}
The detailed derivation of (\ref{eq:5b}) can be found in Appendix~\ref{proof_areaRF}.

Next, let $\Phi(l,w)$ be a point process for the centers of buildings with the same length $l$ and width $w$. Since $\Phi(l,w)$ is a subset of the point process of the centers  of all buildings $\left\{C\right\}$, it is thus a PPP with density $\lambda_{l,w}=\lambda f_L(l)dlf_W(w)dw$~\cite{Bai2014}. Denote by $K_{RF_1}(l,w)$ the number of buildings falling in $S_{FF'I'I}$ with their centers belonging to $\Phi(l,w)$. Consequently, $K_{RF_1}(l,w)$ is a Poisson variable with expectation  
\begin{equation}\label{eq:6}\begin{align}
\mathbb{E}[K_{RF_1}(l,w)]&=\lambda_{l,w}\cdot S_{FF'I'I}\nonumber\\
&=\lambda_{l,w}\frac{l c^2\tau \Delta \tau}{2\sqrt{c^2\tau^2-D^2\cos^2\theta}}.
\end{align}\end{equation}
Note that $K_{RF_1}(l,w)$ is an independent Poisson random variable for different values of $l$ and $w$. Thus, by the superposition property of Poisson random variables~\cite{Stoyan1995}, the total number of buildings falling within $S_{FF'I'I}$ is $K_{RF1}=\sum_{l,w}K_{RF_1}(l,w)$. The expectation of $K_{RF_1}$ is given by
\begin{equation}\begin{align}\label{eq:7a}
~\mathbb{E}[K_{RF_1}]&=\int_L\int_W \frac{\lambda l c^2 \tau \Delta \tau}{2\sqrt{c^2\tau^2-D^2\cos^2\theta}} f_W(w)\,f_L(l)\,dwdl\nonumber\\
&=\lambda\frac{\mathbb{E}[l]  c^2 \tau \Delta \tau}{2\sqrt{c^2\tau^2-D^2\cos^2\theta}}.
\end{align}\end{equation}

Finally, on the basis of (\ref{eq:7a}) and by recalling the definition of $f_{RF_1}(\tau|\theta)$ in (\ref{eq:5}), we have the following theorem.
\begin{theorem}The density of the sub-event that there is a building with orientation $\theta$ to generate a first-order reflection path with delay $\tau$ and its reflection point located at $R_1$, is given by
\begin{equation}\begin{align}\label{eq:7b}
f_{RF_1}(\tau|\theta)=\lambda\frac{\mathbb{E}[l] c\tau}{2\sqrt{c^2\tau^2-D^2\cos^2\theta}}.
\end{align}\end{equation}
\end{theorem}
The proof of Theorem 1 can be found in Appendix~\ref{proof_Theorem1}.
\subsection{Derivation of $f_{NB_1}(\tau|\theta)$}\label{sec:UNOBSTRUCTED}\label{sub:blockage}
\begin{figure}[t]
\begin{center}
\vspace{-2.5em}
\hspace{-4em}
\includegraphics[scale=1.05]{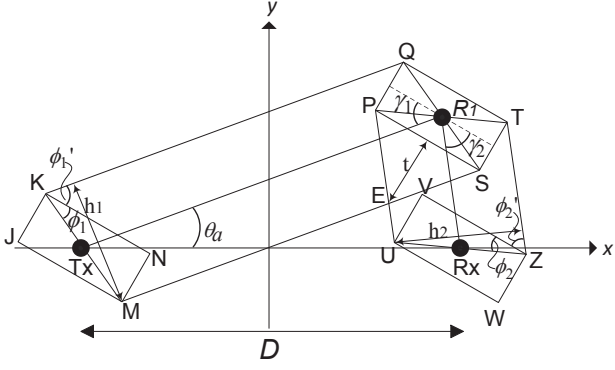}
\hspace{-4em}
\vspace{-1.5em}
\caption{Geometry of the blockage region.}
\vspace{-1em}
\label{fig:3}
\end{center}\end{figure}
Next, let us move our focus to the derivation of $f_{NB_1}(\tau|\theta)$ in (\ref{eq:4}). Suppose that there is a properly located building that can potentially generate a first-order reflection path with its reflection point being at $R_1$. As illustrated in Fig.~\ref{fig:3}, this potential reflection path consists of two line segments, i.e., Tx$R_1$ and $R_1$Rx. It is therefore easy to see that this reflection path can practically exist only if neither of the two line segments, Tx$R_1$ and $R_1$Rx, is blocked by other buildings. Following a similar argument as the blockage analysis of a LoS link in~\cite{Bai2014}, we can conclude that, the probability that the line segment  Tx$R_1$ is not blocked by buildings with length $l$ and width $w$ is equal to the probability that there is no such buildings whose centers lie within the region $JKQTSM$ as illustrated in Fig.~\ref{fig:3}. Similarly, the blockage region for the line segment $R_1$Rx is the region $UPQTZW$. In summary, the probability $f_{NB_1}(\tau|\theta)$ contributed by buildings with length $l$ and width $w$ is equal to the probability that there are no such buildings whose centers lie within the blockage region $JKQTZWUEM$. This blockage region is of a non-convex and irregular shape, whose area is given in the following lemma.

\begin{lemma} A simple yet accurate approximation for the area of the blockage region $JKQTZWUEM$, as illustrated in Fig.~\ref{fig:3}, is given by
\begin{equation}\label{eq:13}\begin{align}
S_{JKQTZWUEM}\approx&~l\sqrt{c^2\tau^2-D^2\cos^2\theta}+wD\left|cos\theta\right|+lw\nonumber\\
&-\frac{l(c\tau-D)}{4}-\frac{l^2\sqrt{c^2\tau^2-D^2}}{8D}.\end{align}\end{equation}
\end{lemma}
The detailed derivation of (\ref{eq:13}) can be found in Appendix~\ref{proof_Lemma2}.

Denote by $K_{NB_1}(l,w)$ the number of buildings with length $l$ and width $w$ whose centers fall in the blockage region $JKQTZWUEM$. Again, we can see that $K_{NB_1}(l,w)$ is a Poisson random variable with expectation
\begin{equation}\label{eq:14}\begin{flalign}
&\mathbb{E}[K_{NB_1}(l,w)]=\lambda_{l,w} \cdot S_{JKQTZWUEM}\nonumber\\
&=\lambda_{l,w}\biggl(l\sqrt{c^2\tau^2-D^2\cos^2\theta}+wD\left|cos\theta\right|+lw-\frac{l(c\tau-D)}{4}\nonumber\\
&~~-\frac{l^2\sqrt{c^2\tau^2-D^2}}{8D}\biggr).
\end{flalign}\end{equation}
With  $K_{NB_1}(l,w)$ being an independent Poisson random variable for a specific value of $l$ and $w$, the total number of buildings falling within the blockage region $S_{JKQTZWUEM}$ is $K_{NB_1}=\sum_{l,w}K_{NB_1}(l,w)$, whose mean value can be calculated as
\begin{equation}\label{eq:15}\begin{flalign}
&\mathbb{E}[K_{NB_1}]=\int_L\int_W \mathbb{E}[K_{NB_1}(l,w)] f_W(w)\,f_L(l)\,dwdl \nonumber\\
&=\lambda\biggl(\mathbb{E}[l]\sqrt{c^2\tau^2-D^2\cos^2\theta}+\mathbb{E}[w]D\left|cos\theta\right|+\mathbb{E}[l]\mathbb{E}[w]\nonumber\\
&~~-\frac{\mathbb{E}[l](c\tau-D)}{4}-\frac{\mathbb{E}[l]^2\sqrt{c^2\tau^2-D^2}}{8D}\biggr).
\end{flalign}\end{equation}
Consequently, on the basis of (\ref{eq:15}), we have the following theorem.
\begin{theorem} Conditioned on that all buildings have orientation $\theta$, the probability that a first-order reflection path with propagation delay $\tau$ and  reflection point located at $R_1$, is not blocked by other buildings is given by
\begin{equation}\label{eq:16}\begin{align}
f_{NB_1}(\tau|\theta)&=\exp\biggl(-\lambda\Bigl(\mathbb{E}[l]\sqrt{c^2\tau^2-D^2\cos^2\theta}\nonumber\\
&~~+\mathbb{E}[w]D\left|cos\theta\right|+\mathbb{E}[l]\mathbb{E}[w]-\frac{\mathbb{E}[l](c\tau-D)}{4}\nonumber\\
&~~-\frac{\mathbb{E}[l]^2\sqrt{c^2\tau^2-D^2}}{8D}\Bigr)\biggr).
\end{align}\end{equation}\end{theorem}
The proof of Theorem 2 is given in Appendix~\ref{proof_Theorem2}.
\subsection{Derivation of $f_{SR}(\tau)$}\label{sec:PROBABILITYSR}
Combining the results derived in the previous two subsections, we are now ready to present the closed-form expression for $f_{SR}(\tau)$. Recalling the definition of a first-order reflection path in (\ref{eq:4}), we can rewrite the function $f_{SR_1}(\tau|\theta)$ by substituting (\ref{eq:7b}) and (\ref{eq:16}) into (\ref{eq:4}) as  
\begin{equation}\label{eq:17}\begin{flalign}
&f_{SR_1}(\tau|\theta)\approx\frac{\lambda\mathbb{E}[l] c\tau}{2\sqrt{c^2\tau^2-D^2\cos^2\theta}} \cdot\nonumber\\
&\exp\biggl(-\lambda\Bigl(\mathbb{E}[l]\sqrt{c^2\tau^2-D^2\cos^2\theta}+\mathbb{E}[w]D\left|cos\theta\right|\nonumber\\
&+\mathbb{E}[l]\mathbb{E}[w]-\frac{\mathbb{E}[l](c\tau-D)}{4}-\frac{\mathbb{E}[l]^2\sqrt{c^2\tau^2-D^2}}{8D}\Bigr)\biggr).\end{flalign}\end{equation}
After similar derivations as those for (\ref{eq:17}), we can obtain
\begin{equation}\label{eq:18a}\begin{flalign}
&f_{SR_2}(\tau|\theta)\approx\frac{\lambda \mathbb{E}[w] c\tau}{2\sqrt{c^2\tau^2-D^2\cos^2\theta}} \cdot\nonumber\\
&\exp\biggl(-\lambda\Bigl(\mathbb{E}[w]\sqrt{c^2\tau^2-D^2\cos^2\theta}+\mathbb{E}[l]D\left|cos\theta\right|\nonumber\\
&+\mathbb{E}[w]\mathbb{E}[l]-\frac{\mathbb{E}[w](c\tau-D)}{4}-\frac{\mathbb{E}[w]^2\sqrt{c^2\tau^2-D^2}}{8D}\Bigr)\biggr),\end{flalign}\end{equation}
\begin{equation}\begin{align}\label{eq:18b}
f_{SR_3}(\tau|\theta)&=f_{SR_1}(\tau|\theta),
\end{align}\end{equation}
and
\begin{equation}\begin{align}\label{eq:18c}
f_{SR_4}(\tau|\theta)&=f_{SR_2}(\tau|\theta).
\end{align}\end{equation}
Therefore, we have
\begin{equation}\label{eq:19a}\begin{flalign}
&f_{SR}(\tau|\theta)=\sum_{i=1}^{4} f_{SR_i}(\tau|\theta)\nonumber\\
&\approx\frac{\lambda \mathbb{E}[l] c\tau}{\sqrt{c^2\tau^2-D^2\cos^2\theta}} \cdot\nonumber\\
&\exp\biggl(-\lambda\Bigl(\mathbb{E}[l]\sqrt{c^2\tau^2-D^2\cos^2\theta}+\mathbb{E}[w]D\left|cos\theta\right|\nonumber\\
&+\mathbb{E}[l]\mathbb{E}[w]-\frac{\mathbb{E}[l](c\tau-D)}{4}-\frac{\mathbb{E}[l]^2\sqrt{c^2\tau^2-D^2}}{8D}\Bigr)\biggr)\nonumber\\
&+\frac{\lambda \mathbb{E}[w] c\tau}{\sqrt{c^2\tau^2-D^2\cos^2\theta}} \cdot\nonumber\\
&\exp\biggl(-\lambda\Bigl(\mathbb{E}[w]\sqrt{c^2\tau^2-D^2\cos^2\theta}+\mathbb{E}[l]D\left|cos\theta\right|\nonumber\\
&+\mathbb{E}[w]\mathbb{E}[l]-\frac{\mathbb{E}[w](c\tau-D)}{4}-\frac{\mathbb{E}[w]^2\sqrt{c^2\tau^2-D^2}}{8D}\Bigr)\biggr).\end{flalign}\end{equation}

Note that the expression (\ref{eq:19a}) is derived based on the assumption of $\theta \in (\pi/2,\pi]$ defined in Section III-A. For the case when $\theta \in (0,\pi/2]$, as illustrated in Fig.~$\ref{fig:ellipse4Q}$, we can see that the angle between the $x$-axis and the tangent line of the ellipse across point $R_1$ is $\varphi=\theta+\frac{\pi}{2}$. Hence following a similar derivation as (\ref{derive_RF2}), the slope of the tangent line across point $R_1$ in this case is given by
\begin{equation}\label{eq:19a1}
\tan \varphi=\tan \left(\theta+\frac{\pi}{2}\right)= \frac{dy}{dx}=\frac{-n^2x}{m^2y}.
\end{equation}
Next, to obtain the function $f_{SR}(\tau|\theta)$ for the case when $\theta \in (0,\pi/2]$, we can follow the similar derivations as those from (\ref{eq:3a}) to (\ref{eq:19a}), except that we replace the expression for the slope of the tangent line in (\ref{derive_RF2}) with (\ref{eq:19a1}) and exchange the building parameter $l$ and $w$. Thus, we have
\begin{equation}\label{eq:19b}\begin{flalign}
&f_{SR}(\tau|\theta)=\sum_{i=1}^{4} f_{SR_i}(\tau|\theta)\nonumber\\
&\approx\frac{\lambda \mathbb{E}[w] c\tau}{\sqrt{c^2\tau^2-D^2\cos^2\varphi}} \cdot\nonumber\\
&\exp\biggl(-\lambda\Bigl(\mathbb{E}[w]\sqrt{c^2\tau^2-D^2\cos^2\varphi}+\mathbb{E}[l]D\left|cos\varphi\right|\nonumber\\
&+\mathbb{E}[w]\mathbb{E}[l]-\frac{\mathbb{E}[w](c\tau-D)}{4}-\frac{\mathbb{E}[w]^2\sqrt{c^2\tau^2-D^2}}{8D}\Bigr)\biggr)\nonumber\\
&+\frac{\lambda \mathbb{E}[l] c\tau}{\sqrt{c^2\tau^2-D^2\cos^2\varphi}} \cdot\nonumber\\
&\exp\biggl(-\lambda\Bigl(\mathbb{E}[l]\sqrt{c^2\tau^2-D^2\cos^2\varphi}+\mathbb{E}[w]D\left|cos\varphi\right|\nonumber\\
&+\mathbb{E}[l]\mathbb{E}[w]-\frac{\mathbb{E}[l](c\tau-D)}{4}-\frac{\mathbb{E}[l]^2\sqrt{c^2\tau^2-D^2}}{8D}\Bigr)\biggr)\end{flalign}\end{equation}
where $\varphi=\theta+\frac{\pi}{2}$.

Finally, by substituting (\ref{eq:19a}) and (\ref{eq:19b}) into (\ref{eq:3}), we can obtain the closed-form expression for $f_{SR}(\tau)$ as  
\begin{equation}\label{eq:20}\begin{align}
f_{SR}(\tau)&\approx\zeta_1\biggl(\frac{8\eta}{\pi}(1+\eta\beta_1)tanh^{-1}{\left(\frac{\pi}{8\eta^2}\right)}-\beta_1\biggr)\nonumber\\
&~~+\zeta_2\biggl(\frac{8\eta}{\pi}(1+\eta\beta_2)tanh^{-1}{\left(\frac{\pi}{8\eta^2}\right)}-\beta_2\biggr)
\end{align}\end{equation}
where
\begin{equation}\label{eq:21}\begin{align}
\zeta_1&=\lambda\mathbb{E}[l]a\exp\biggl(\lambda\mathbb{E}[l]\left(\frac{D(a-1)}{4}-D\eta-\mathbb{E}[w]\right)\biggr)\nonumber\\
&\times\exp\biggl(\frac{\lambda\mathbb{E}[l]^2\sqrt{a^2-1}}{8}+\frac{\lambda\mathbb{E}[w]D}{\sqrt{a^2-1}-a}(a-\eta)\biggr)\nonumber\\
\zeta_2&=\lambda\mathbb{E}[w]a\exp\biggl(\lambda\mathbb{E}[w]\left(\frac{D(a-1)}{4}-D\eta-\mathbb{E}[l]\right)\biggr)\nonumber\\
&\times\exp\biggl(\frac{\lambda\mathbb{E}[w]^2\sqrt{a^2-1}}{8}+\frac{\lambda\mathbb{E}[l]D}{\sqrt{a^2-1}-a}(a-\eta)\biggr)\nonumber\\
\beta_1&=\lambda D \biggl(\mathbb{E}[l]+\frac{\mathbb{E}[w]}{\sqrt{a^2-1}-a}\biggr)\nonumber\\
\beta_2&=\lambda D \biggl(\mathbb{E}[w]+\frac{\mathbb{E}[l]}{\sqrt{a^2-1}-a}\biggr)\nonumber\\
\eta&=\sqrt{a^2-(1/2)}\nonumber\\
a&=\frac{L_r}{D}=\frac{c \tau}{D}.\nonumber\end{align}\end{equation}
The detailed derivation of (\ref{eq:20}) can be found in Appendix~\ref{proof_closedform}.
\subsection{PDP and Average Number of the First-Order Reflection Paths}
We now return to the PDP contributed by all the first-order reflection paths as defined in (\ref{eq:1}). Substituting (\ref{eq:20}) into (\ref{eq:1}), we have
\begin{equation}\label{eq:20a}\begin{align}
P(\tau)\approx&\rho(\tau)\cdot \Biggl(\zeta_1\biggl(\frac{8\eta}{\pi}(1+\eta\beta_1)tanh^{-1}{\left(\frac{\pi}{8\eta^2}\right)}-\beta_1\biggr)\nonumber\\
&~~+\zeta_2\biggl(\frac{8\eta}{\pi}(1+\eta\beta_2)tanh^{-1}{\left(\frac{\pi}{8\eta^2}\right)}-\beta_2\biggr)\Biggr).
\end{align}\end{equation}
In brief, the values of $\zeta_1, \zeta_2, \beta_1$ and $\beta_2$ in (\ref{eq:20a}) all depend on the average dimensions of the buildings $\mathbb{E}[l]$ and $\mathbb{E}[w]$, and the density of the buildings $\lambda$. Thus we can conclude that the behavior of PDP contributed by all the first order reflection paths varies with these environment parameters. We will further discuss the effect of these parameters in the next section. 

Note that, although the resultant expression (\ref{eq:20a}) is a bit complicated, it is much more efficient to be used for analyzing the PDP compared to numerical simulations. Therefore, the derived expression can effectively facilitate the system modeling and greatly reduce the time for numerical simulations, which is the key contribution of our work.

On top of (\ref{eq:20a}), we can obtain the total path loss of a link by combining the average path loss of a LoS component and the average path loss of the first-order reflection paths, which are denoted by $\mathbb{E}[P_{LoS}]$ and $\mathbb{E}[P_{Ref}]$, respectively. The total path loss of a link is given by
\begin{equation}\label{path_loss_total}
\mathbb{E}[P_t]=\mathbb{E}[P_{LoS}]+\mathbb{E}[P_{Ref}].
\end{equation}
In (\ref{path_loss_total}), the average path loss of the LoS component is given by \cite{Bai2014}
\begin{equation}\begin{align}\label{path_loss_LoS}
\mathbb{E}[P_{LoS}]=&\frac{1}{\left(4\pi f \tau_0\right)^2}\cdot\nonumber\\
&\exp\left(-\frac{2\lambda(\mathbb{E}[L]+\mathbb{E}[W])D}{\pi}-\lambda\mathbb{E}[L]\mathbb{E}[W]\right)
\end{align}\end{equation}
and the average path loss of the first-order reflection paths can be obtained by
\begin{equation}\label{path_loss_reflection}
\mathbb{E}[P_{Ref}]=\int_{\tau_0}^{\tau_{max}}P(\tau) \,d\tau.
\end{equation}
In (\ref{path_loss_reflection}), $\tau_0$ is the minimum propagation delay of all the first-order reflection paths, which is lower bounded by the propagation delay of the LoS path, i.e., we can set $\tau_0 = D/c$. Similarly, $\tau_{max}$ is the maximum propagation delay of all first-order reflection paths, which should be set as $\tau_{max} = +\infty$ theoretically.

As a by-product, the average number of the first-order reflection paths, which is denoted by $\mathbb{E}[f_{SR}]$, can be derived from (\ref{eq:20}) as
\begin{equation}\label{eq:average_no of path1}
\mathbb{E}[f_{SR}]=\int_{\tau_0}^{\tau_{max}}f_{SR}(\tau) \,d\tau.
\end{equation}

\section{Numerical Results and Discussions}\label{sec:numerical}
In this section, we will present some numerical examples to validate the accuracy of the proposed analytical model and discuss the impact of environment features, e.g., building dimensions and density, on the PDP and the average number of the first-order reflection paths.
\subsection{Validation of the Blockage Area Approximation (\ref{eq:13})}
\begin{figure}[t]\centering
\vspace{-1.5em}
\includegraphics[scale=0.5]{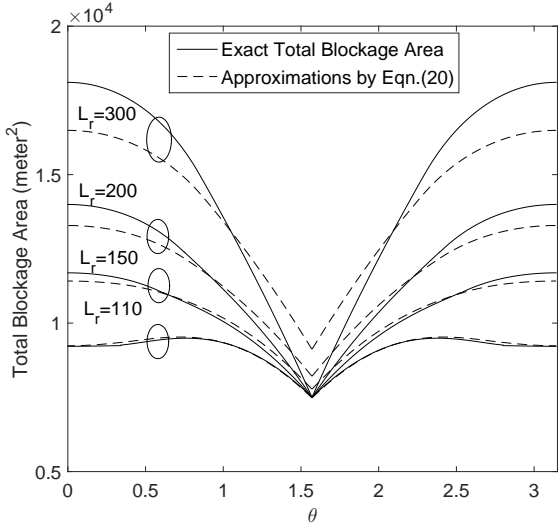}
\caption{Comparison between the exact and approximated total blockage area for the orientation of buildings $\theta\in(0,\pi)$.}
\label{fig:AreaSPE}
\end{figure} 
First, let us verify the accuracy of the blockage area approximation in (\ref{eq:13}). In Fig.~\ref{fig:AreaSPE}, both the exact total blockage area and the approximation (\ref{eq:13}) are plotted as functions of the building orientation $\theta$ when $D=100$ m, $l=55$ m and $w=50$ m. After comparison, we observe that the approximation  involved in (\ref{eq:13}) is very accurate for small values of the path length $L_r$ in the whole interval $\theta\in(0,\pi]$. Though the gap between the exact and approximation curves becomes visible as the value of $L_r$ increases, we observe that by using (\ref{eq:13}), we will underestimate the blockage area when the value of $\theta$ is close to $0$ or $\pi$, and overestimate the blockage area when the value of $\theta$ is close to $\pi/2$. Recall from (\ref{eq:3}) that the function $f_{SR}(\tau)$ involves the integration with respect to $\theta$ from $0$ to $\pi$. Intuitively, after performing the integration, we can expect that the overestimated and underestimated areas will cancel each other to a certain extent and reduce the approximation error. We will further verify the accuracy of this approximation in the next subsection. 
\subsection{Effects of Environment Parameters on PDP}\label{effect}
\begin{figure}[t]\centering
\vspace{-1.5em}
\includegraphics[width=7.5cm,height=7.5cm]{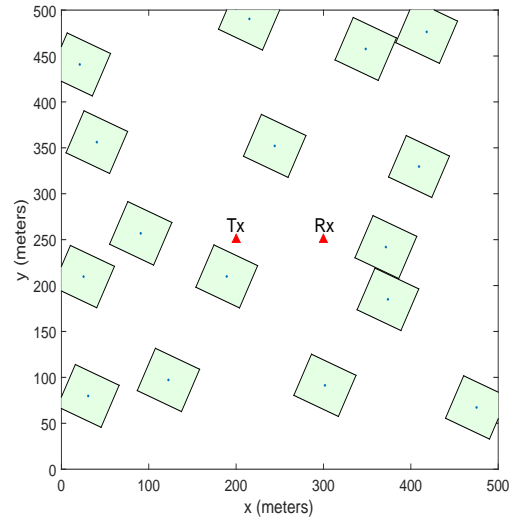}
\caption{An example of a point-to-point link with distance $D=100$ m. This link is surrounded by randomly located buildings with $l\in[54,56]$ and $w\in[49,51]$.}
\vspace{-1em}
\label{RandomBuildings}
\end{figure}
We now set about validating our analytical results and discussing the effect of environment features on the PDP contributed by all the first-order reflection paths. We consider a point-to-point link with distance $D=100$ m and operating frequency $f=73$~GHz. Both the transmitter and receiver are equipped with a single omnidirectional antenna with unit gain. Such a link is surrounded by randomly distributed and identically orientated buildings. The lengths and widths of the buildings are uniformly distributed in certain intervals. The center locations of all the buildings in this area form a homogeneous PPP with density $\lambda$. For the ease of subsequent discussions, we denote $\Phi$ as the fraction of land covered by buildings in the investigated area, i.e., $\Phi=1-\exp(-\lambda\mathbb{E}[l]\mathbb{E}[w])$\cite{Bai2014}, which we refer to as the covered ratio. 

Given the detailed parameter setting in the above system, the PDP contributed by all the first-order reflection paths can be analytically obtained by (\ref{eq:20a}). Besides, we also provide the simulation results of the PDP contributed by all the first-order reflection paths in the following way. We consider a square area\footnote{An infinite area should be considered here theoretically, but this is impossible in simulations. However, we argue that only considering such a finite square area is sufficient, as the PDP is mainly contributed by short reflection paths and the received power through those long reflection paths is very insignificant after much severer propagation loss. We have numerically verified this via varying the size of this square area.} of $500\times500$ m$^2$ as illustrated in Fig.~\ref{RandomBuildings}, where the transmitter and receiver are located at $(200$m$,250$m$)$ and $(300$m$,250$m$)$, respectively. In each channel realization, we first randomly generate buildings within this square area based on the given building parameters. Then we follow the ray-tracing principle and the law of reflection to check if every single side wall of each building is able to generate a first-order reflection path between the transmitter and receiver. The blockage status of the potential first-order reflection paths is then further verified. We set the reflection loss $\sigma=3$~dB. Each first-order reflection path is specified by its time delay, and the corresponding path loss can be calculated consequently. After simulating a sufficiently large number of channel realizations, we sort all the collected first-order reflection paths by their time delays and classify them into pre-partitioned time delay intervals. Finally, the sum path loss of all the first-order reflection paths in each interval is divided by the product of the interval width and the total number of channel realizations. The resultant ratios are referred to as the simulated PDP contributed by all the first-order reflection paths. 

\begin{figure}[tbp]
\centering
\subfloat[]{\label{fig:Result2a}\includegraphics[width=9cm,height=7.0cm]{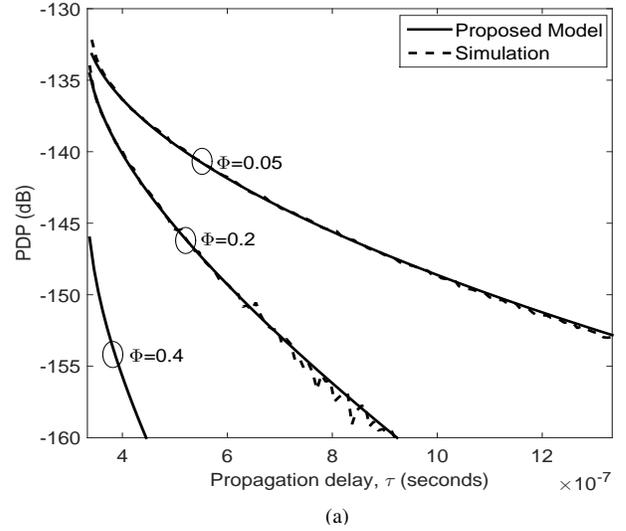}}\vspace{-1.2em}\\
\subfloat[]{\label{fig:Result2b}\includegraphics[width=9cm,height=7.0cm]{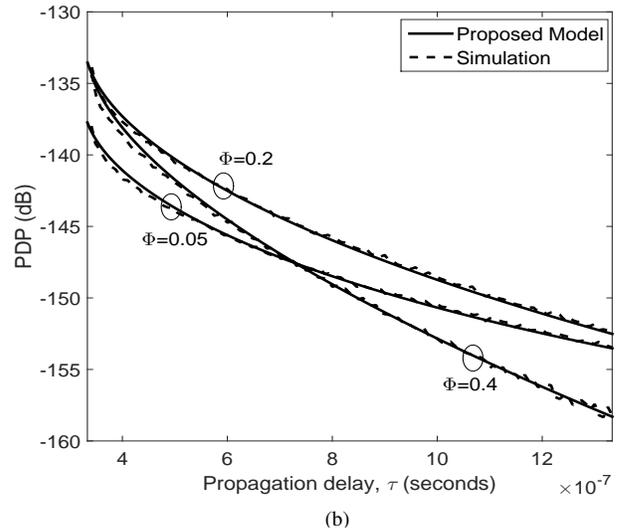}}\vspace{-1.2em}\\
\subfloat[]{\label{fig:Result2c}\includegraphics[width=9cm,height=7.0cm]{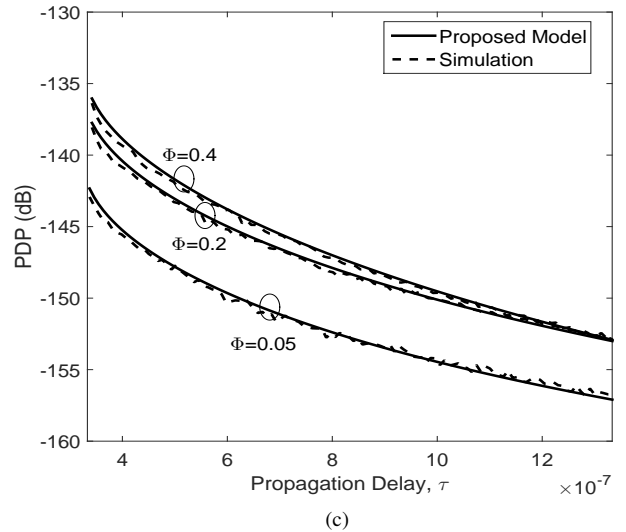}}
\caption{Comparison between analytical and simulated PDP. The dimensions of buildings are (a) $l\in[9$~m,$11$~m$]$ and $w\in[9$~m,$11$~m$]$, (b) $l\in[54$~m,$56$~m$]$ and $w\in[49$~m,$51$~m$]$, (c) $l\in[149$~m,$151$~m$]$ and $w\in[149$~m,$151$~m$]$.}
\label{fig:Result2Main}
\end{figure}
Fig.~\ref{fig:Result2Main} compares the analytical and simulated PDPs under various building environments. In each sub-figure of Fig.~\ref{fig:Result2Main}, three values of the covered ratio, i.e., $\Phi=0.05, 0.2$ and $0.4$,  are considered, which represent sparse, medium and dense building environments, respectively. In addition, three different pairs of length and width distributions are considered in Figs.~\ref{fig:Result2Main}\subref{fig:Result2a},~\ref{fig:Result2Main}\subref{fig:Result2b} and~\ref{fig:Result2Main}\subref{fig:Result2c} to observe the effect of the building dimensions on PDP. Specifically, we assume the uniform distributions of $l\in[9$~m,$11$~m$]$ and $w\in[9$~m,$11$~m$]$ in Fig.~\ref{fig:Result2Main}\subref{fig:Result2a}; those of $l\in[54$~m,$56$~m$]$ and $w\in[49$~m,$51$~m$]$ in Fig.~\ref{fig:Result2Main}\subref{fig:Result2b}, and those of $l\in[149$~m,$151$~m$]$ and $w\in[149$~m,$151$~m$]$ in Fig.~\ref{fig:Result2Main}\subref{fig:Result2c}, which represent small, medium and large sizes of buildings, respectively. From these figures\footnote{Note that the simulated PDP for the case of $\Phi=0.4$ is not presented in Fig.~\ref{fig:Result2Main}\subref{fig:Result2a}, because the related simulation is measured in months and very time-consuming.}, we can make the following observations.
\begin{itemize}
\item The analytical PDP derived in (\ref{eq:20a}) is numerically very accurate in all scenarios, which indicates that the error incurred by  the blockage area approximation in (\ref{eq:13}) is negligible. Note that the calculation of equation (\ref{eq:20a}) only takes less than a second, whereas the time of obtaining those simulation curves in Fig.~\ref{fig:Result2Main} is measured in days. Therefore, though the expression in equation (\ref{eq:20a}) looks complex, it provides a very efficient way to analyze the PDP.
\item Given the distributions of building dimensions, there exists the most
preferable value of covered ratio between $0$ to $1$ that leads to the maximum PDP curve. This is straightforward as in one extreme of $\Phi=0$, there is no building to generate reflections, whereas in another extreme of $\Phi=1$, the land is all covered by buildings and so all potential reflection paths will be blocked. Therefore, the value of $\Phi$ that leads to the maximum PDP curve must fall between $0$ and $1$. 
\item The most preferable value of $\Phi$ increases with the sizes of buildings, i.e., the length and width. For example, with small buildings in Fig.~\ref{fig:Result2Main}\subref{fig:Result2a}, the PDP curve for $\Phi=0.05$ is the highest among the three PDP curves. As the building sizes increase, in Fig.~\ref{fig:Result2Main}\subref{fig:Result2b} and Fig.~\ref{fig:Result2Main}\subref{fig:Result2c}, the highest ones are achieved when $\Phi=0.2$ and $0.4$, respectively. Thus we can have this observation qualitatively.
\item The PDP curves decrease faster with the time delay $\tau$ when the building size is smaller or the covered ratio is larger. For example, by comparing the PDP curves for the case of $\Phi=0.05$ in all subfigures, it is seen that the slope of the PDP curve in Fig.~\ref{fig:Result2Main}\subref{fig:Result2a}, i.e., for small buildings, is sharper than the other PDP curves. In addition, it is also seen from Fig.~\ref{fig:Result2Main}\subref{fig:Result2a} that the PDP curve for $\Phi=0.4$ has the sharpest slope among other PDP curves. This is because the existence of dense buildings in small dimensions will significantly increase the blockage probability of the first-order reflection paths.
\end{itemize}   

\begin{figure}[t!]\centering
\includegraphics[width=9cm,height=7.5cm]{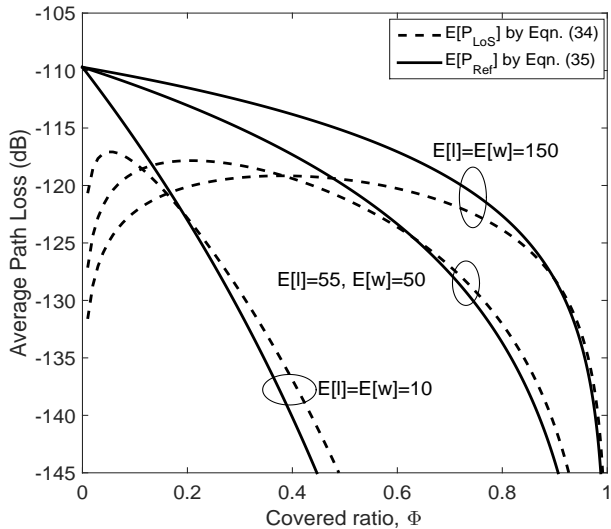}
\caption{Effect of the covered ratio on the average path loss.}
\label{fig:Result3a}
\end{figure}

\begin{figure}[t!]\centering
\includegraphics[width=9cm,height=7.5cm]{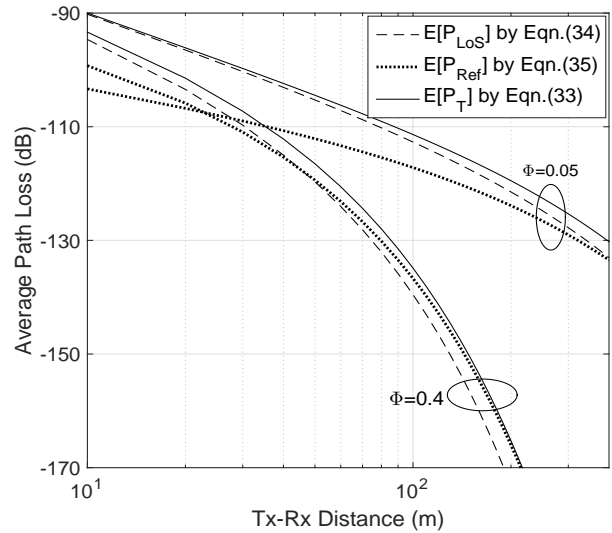}
\footnotesize\caption{Effect of Tx-Rx distance on the average path loss.}
\label{fig9}
\end{figure}

Next, we compare the average path loss contributed by the first-order reflection paths in (\ref{path_loss_reflection}) with the average path loss contributed by the LoS path in (\ref{path_loss_LoS}). We use the same system setting and building parameters as in Fig.~\ref{fig:Result2Main}. Thus, we have the lower bound of the time delay in (\ref{path_loss_reflection}) $\tau_0=D/c=0.33\mu s$. To facilitate the numerical integration in (\ref{path_loss_reflection}), we set the upper bound in (\ref{path_loss_reflection}) to be a sufficiently large but finite value\footnote{Theoretically, $\tau_{max}$ should be $+\infty$, but it is impossible in the numerical integration. However, the finite value of $\tau_{max}$ is a reasonable setting because in practice, the reflected waves with long delays contribute to insignificant received powers, which can be ignored. A similar treatment has also been adopted in~\cite{Meijerink2014,Liberti1999}.}. Fig.~\ref{fig:Result3a} plots the average path loss contributed by the first-order reflection paths and LoS path as functions of covered ratio $\Phi$. From this figure, we can make the following observations.
\begin{itemize}
\item Different from the monotonously decreasing LoS curves, the average path loss contributed by the first-order reflection paths exhibits as a concave function of the covered ratio. The optimal scenario that leads to the minimum absolute value of the average path loss occurs when $\Phi$ is between $0$ and $1$. 
\item In systems with $\Phi$ higher than this optimal scenario, the average path loss contributed by all the first-order reflection paths becomes comparable with that contributed by the LoS path, indicating that the reflection paths are not ignorable in these cases.
\item The value of $\Phi$ corresponding to this optimal scenario increases as the size of buildings increases. For example, it is seen from Fig.~\ref{fig:Result3a} that the most preferable value of covered ratio for the small, medium and large buildings, are respectively around, $\Phi=0.05$, $\Phi=0.2$ and $\Phi=0.4$.   
\end{itemize}
Then, in Fig.~\ref{fig9}, we plot the total path loss, the average path loss of the LoS and the average path loss of the first-order reflection paths, which are respectively obtained by (\ref{path_loss_total}), (\ref{path_loss_LoS}) and (\ref{path_loss_reflection}), as functions of Tx-Rx distance $D$. We use the average sizes of building $\mathbb{E}[L]=\mathbb{E}[W]=10$ m, which represent small sizes of buildings. Two values of covered ratio $\Phi$ are considered, i.e., $\Phi=0.05$ and $0.4$. From this figure, we can make the following observations.
\begin{itemize}
\item In the system with a low value of $\Phi$, the total path loss is dominated by the contribution of the LoS path, especially at the short Tx-Rx distance.
\item Given the average sizes of buildings, there exist an optimum value of $D$ at which both the LoS and the first-order reflection paths contribute to the same average path loss. 
\item The value of the optimal $D$ decreases as the covered ratio $\Phi$ increases. From Fig.~\ref{fig9}, it is seen that the values of the optimal $D$ for the covered ratio $\Phi=0.05$ and $\Phi=0.4$, are respectively, $390$ m and $30$ m.
\end{itemize}
The same observations are found for the cases of medium and large sizes of buildings.
\subsection{Average Number of the First-Order Reflection Paths}
\begin{figure}[tbp]\centering
\includegraphics[width=9cm,height=7.5cm]{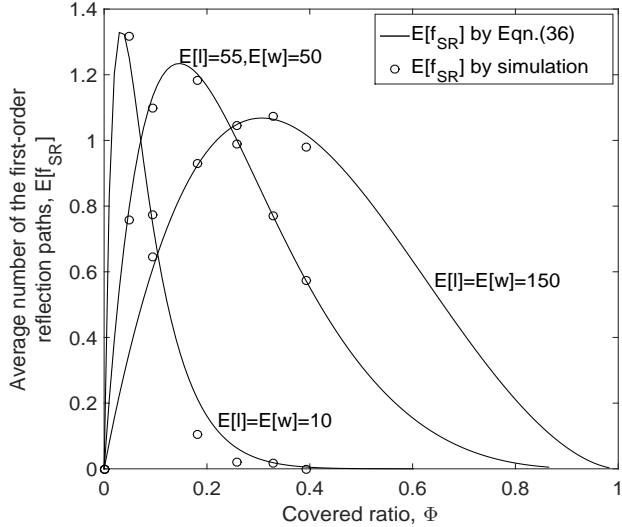}
\caption{Effect of the covered ratio on the average number of the first-order reflection paths.}
\label{fig:3b}
\end{figure}

Finally, we discuss the effect of the environment features on the average number of the first-order reflection paths $\mathbb{E}[f_{SR}]$. Fig.~\ref{fig:3b} plots both the semi-analytical and simulated $\mathbb{E}[f_{SR}]$ as functions of covered ratio $\Phi$, where the solid curves are calculated based on (\ref{eq:average_no of path1}) and numerical integration, and the black circles are simulation results obtained by the same simulation process as in Section \ref{effect}. The following observations are made from Fig.~\ref{fig:3b}. 
\begin{itemize}
\item The semi-analytical $\mathbb{E}[f_{SR}]$ in (\ref{eq:average_no of path1}) is accurate in all scenarios. Although we are unable to derive the closed expression for $\mathbb{E}[f_{SR}]$, the semi-analytical expression in (\ref{eq:average_no of path1}) provides an efficient way to analyze the contribution of the first-order reflection paths.  
\item All $\mathbb{E}[f_{SR}]$ curves exhibit a unimodal property in the range $\Phi\in[0,1]$. These curves first increase monotonically as $\Phi$ increases from zero. After achieving their maximum points, these curves decrease with the increasing value of $\Phi$.   
\item There exists a peak point for each $\mathbb{E}[f_{SR}]$ curve, which indicates the most preferable value of the covered ratio that leads to the maximum $\mathbb{E}[f_{SR}]$. The most preferable value of the covered ratio $\Phi$ increases with the size of buildings. This observation is consistent with the behavior of PDP in Section \ref{effect}.
\end{itemize}
\section{Conclusion}
In this paper, we have developed a stochastic channel model for point-to-point MMW systems in outdoor environments. Our channel model incorporates the environment features to capture the effects of the first-order reflection paths generated by randomly distributed buildings. We developed an approximate but accurate closed-form expression for the PDP contributed by all the first-order reflection paths and derived a semi-analytical expression for the average number of the first-order reflection paths. Numerical results demonstrated that these approximate expressions are very tight under all the considered system settings. Our results illustrate that wireless networks may benefit from buildings in the area of MMW communication links, since the external surfaces of these buildings render reflection paths that can provide a comparable signal power to that of the LoS path. Our results also reveal that there exists an optimal system setting that can contribute to the maximum power of the first-order reflection paths. The findings in this paper can provide useful insights to develop more complex channel models applicable to future MMW systems.     
\appendix
\subsection{Derivation of (\ref{eq:5b})}\label{proof_areaRF}
To formulate $S_{FF'I'I}$ in (\ref{eq:5b}), let us return to (\ref{eq:5a}). After similar derivations as those for (\ref{derive_RF4}) and (\ref{derive_RF5}), except that we replace $\tau$ with $\tau+\Delta \tau$, we have the coordinates of the reflection point $R_1'$, denoted by $(x', y')$, as
\begin{equation}\label{proof_areaRF_51}\begin{align}
x'&=\frac{-c^2(\tau+\Delta\tau)^2\tan \theta}{2\sqrt{c^2(\tau+\Delta\tau)^2\sec^2 \theta-D^2}}\nonumber\\
&=x+\frac{c^2\tau\Delta\tau\tan \theta(2D^2-c^2\tau^2\sec^2\theta)}{2(c^2\tau^2\sec^2 \theta-D^2)^{3/2}},
\end{align}
\end{equation}
and
\begin{equation}\label{proof_areaRF_61}\begin{align}
y'&=\frac{c^2(\tau+\Delta\tau)^2-D^2}{2\sqrt{c^2(\tau+\Delta\tau)^2\sec^2 \theta-D^2}}\nonumber\\
&=y+\frac{c^2\tau\Delta\tau(c^2\tau^2\sec^2-2D^2+D^2\sec^2\theta)}{2(c^2\tau^2\sec^2 \theta-D^2)^{3/2}}.
\end{align}
\end{equation}
Note that the second equalities of (\ref{proof_areaRF_51}) and (\ref{proof_areaRF_61}) are obtained by ignoring all terms that contain $\Delta\tau^2$. 

Next, we obtain the vector and magnitude of the line segment $FF'$, which corresponds to the line segment $RR'$ as illustrated in Fig.~\ref{fig:3abc}(c) as, respectively, 
\begin{equation}\label{proof_areaRF_7}
\begin{split}
\overrightarrow{FF'}&=\left[\begin{array}{ccc} x-x'\\
 y-y'\end{array}\right]=\left[\begin{array}{ccc}
~\frac{c^2\tau\Delta\tau\tan \theta(2D^2-c^2\tau^2\sec^2\theta)}{2(c^2\tau^2\sec^2 \theta-D^2)^{3/2}}\\
~\frac{c^2\tau\Delta\tau(c^2\tau^2\sec^2-2D^2+D^2\sec^2\theta)}{2(c^2\tau^2\sec^2 \theta-D^2)^{3/2}}
\end{array}\right],
\end{split}
\end{equation}
and
\begin{equation}\begin{align}\label{proof_areaRF_12} 
|\overrightarrow{FF'}|&=\sqrt{(x'-x)^2+(y'-y)^2}\nonumber\\
&=\frac{c^2\tau\Delta \tau\sqrt{c^4\tau^4\sec^6\theta+D^4\sec^4\theta-2D^2c^2\tau^2\sec^4\theta}}{2(c^2\tau^2\sec^2\theta-D^2)^{3/2}}.
\end{align}\end{equation}
Meanwhile, the vector and magnitude of the line segment $FI$ are given as
\begin{equation}\label{proof_areaRF_8}
\begin{split}
\overrightarrow{FI} &=\left[\begin{array}{ccc} -l \cos⁡  \theta\\
-l \sin  ⁡ \theta\end{array}\right],
\end{split}
\end{equation}
and
\begin{equation}\label{proof_areaRF_11}
|\overrightarrow{FI}|=\sqrt{l^2\cos^2\theta + l^2\sin^2\theta}=l.
\end{equation}
Denote by $\psi$ the angle between line segments $FF'$ and $FI$, where we have
\begin{equation}\label{proof_areaRF_9}
\cos\psi =\frac{1}{|\overrightarrow{FF'}||\overrightarrow{FI}|}\frac{-lc^2\tau\Delta\tau D^2\sec^2\theta\sin\theta}{2(c^2\tau^2\sec^2\theta-D^2)^{3/2}}.
\end{equation}
Mathematically, the area of the parallelogram $FF'I'I$ is given as 
\begin{equation}\label{proof_areaRF_6a}\begin{align}
S_{FF'I'I}&=|\overrightarrow{FF'}||\overrightarrow{FI}|\sin\psi=|\overrightarrow{FF'}||\overrightarrow{FI}|\sqrt{1-\cos^2\psi}\nonumber\\
&=\sqrt{|\overrightarrow{FF'}|^2|\overrightarrow{FI}|^2-(|\overrightarrow{FF'}||\overrightarrow{FI}|\cos\psi)^2}\nonumber\\
\end{align}\end{equation}
Finally, by substituting (\ref{proof_areaRF_12}),(\ref{proof_areaRF_11}),(\ref{proof_areaRF_9}) into (\ref{proof_areaRF_6a}), we have the expression for $S_{FF'I'I}$ as written in (\ref{eq:5b}).
\subsection{Proof of \textbf{Theorem 1}}\label{proof_Theorem1}
The first-order reflection path with its reflection point at $R_1$ will occur when there is at least one building whose center falls within the $FF'I'I$ region. Thus we have
\begin{equation}\label{eq:8a}\begin{align}
\mathbb{P}(E_1|\mathcal{N}_{R_1})&=\mathbb{P}(K_{RF_1}>0)=1-\mathbb{P}(K_{RF_1}=0)
\end{align}\end{equation}
where\footnote{For a Poisson distributed random variable $x$ with expectation $\mathbb{E}[x]$, we have $\mathbb{P}(x=n)=\frac{\mathbb{E}[x]^n e^{-\mathbb{E}[x]}}{n!}$~\cite{Lennart2000}.}  
\begin{equation}\label{eq:8b}\begin{align}
\mathbb{P}(K_{RF_1}=0)&=\frac{\left(\mathbb{E}[K_{RF_1}]\right)^0}{0!}e^{-\mathbb{E}[K_{RF_1}]}\nonumber\\
&=e^{-\lambda\frac{\mathbb{E}[l] c^2\tau \Delta \tau}{2\sqrt{c^2\tau^2-D^2\cos^2\theta}}}.
\end{align}\end{equation}
Substituting (\ref{eq:5a}) and (\ref{eq:8a}) into (\ref{eq:5}), we have
\begin{equation}\label{eq:8c}\begin{align}
f_{RF_1}(\tau|\theta)&=\displaystyle\lim_{c\Delta \tau\to 0}\frac{1-e^{-\lambda\frac{\mathbb{E}[l] c^2\tau \Delta \tau}{2\sqrt{c^2\tau^2-D^2\cos^2\theta}}}}{c\Delta \tau}.
\end{align}\end{equation}
Finally, using the L'Hospital rule~\cite{Stefan2000}, we can rewrite (\ref{eq:8c}) as
\begin{equation}\label{eq:8d}\begin{align}
f_{RF_1}(\tau|\theta)&=\displaystyle\lim_{c\Delta \tau\to 0}\frac{\lambda\frac{\mathbb{E}[l] c\tau}{2\sqrt{c^2\tau^2-D^2\cos^2\theta}}e^{-\lambda\frac{\mathbb{E}[l] c^2\tau \Delta \tau}{2\sqrt{c^2\tau^2-D^2\cos^2\theta}}}}{1}\nonumber\\
&=\lambda\frac{\mathbb{E}[l] c\tau}{2\sqrt{c^2\tau^2-D^2\cos^2\theta}}. 
\end{align}\end{equation} which completes the proof.\\
\subsection{Derivation of (\ref{eq:13})}\label{proof_Lemma2}
From Fig.~\ref{fig:3}, the area of the blockage region $JKQTZWUEM$ can be express as  
\begin{equation}\begin{align}\label{eq:10}
S_{JKQTZWUEM}&=S_{JKQTSM}+S_{UPQTZW}-S_{PQTS}\nonumber\\
&~~-S_{PSE},
\end{align}\end{equation}
where each term on the right hand-side is interpreted individually as follows. Specifically, the region $JKQTSM$ consists of two right-angled triangles $JKM$ and $QTS$, and a parallelogram $KQSM$ with height denoted by $h_1$ and base $|$Tx$R_1|$, which is the length of the line segment Tx$R_1$. This area is given by
\begin{equation}\label{eq:10a} 
S_{JKQTSM}=\frac{lw}{2}+\frac{lw}{2}+|\text{Tx}R_1|\cdot h_1=lw+|\text{Tx}R_1|\cdot h_1.
\end{equation}
Let us first return to Fig.~\ref{fig:3} and introduce some notations to facilitate the derivation of $h_1$. Denote by $\gamma_1$ the angle between the $l$-side wall of building and the line segment Tx$R_1$. As illustrated in Fig.~\ref{fig:3}, the value of $h_1$ can be obtained by the trigonometric formula as
\begin{equation}\label{proof_h1}\begin{align}
h_1&=\sin(\phi_1+\phi_1')\sqrt{l^2+w^2},
\end{align} 
\end{equation}
where $\phi_1$ is the angle between the $l$-side wall of building and the line segment $KM$ and $\phi_1'=\gamma_1$. Then, by using the trigonometric identity, we have
\begin{equation}\label{proof_h1a}\begin{align} 
h_1&=\left(\sin\gamma_1\cos\phi_1+\cos\gamma_1\sin\phi_1\right)\sqrt{l^2+w^2}\nonumber\\
&=\left(\sin\gamma_1\frac{l}{\sqrt{l^2+w^2}}+\cos\gamma_1\frac{w}{\sqrt{l^2+w^2}}\right)\sqrt{l^2+w^2}.
\end{align} \end{equation}
 
Next, let us derive the expression for $\sin\gamma_1$ and $\cos\gamma_1$. Denote by $\theta_a$ the angle between the $x$-axis and the line segment Tx$R_1$ as illustrated in Fig.~\ref{fig:3}. The slope of the line segment Tx$R_1$ is given by
\begin{equation}\label{proof_h2}
\begin{align}
\tan\theta_a=\frac{y}{x+D/2}.
\end{align} 
\end{equation} 
By recalling the definition of $\theta$ in assumption~\ref{A4}, we have
\begin{equation}\label{proof_h3}
\begin{align}
\gamma_1=\pi-(\theta-\theta_a).
\end{align} 
\end{equation}  
Applying the trigonometric function on both sides of (\ref{proof_h3}), we have
\begin{equation}\label{proof_h4}
\begin{align}
\tan\gamma_1&=\tan(\pi-(\theta-\theta_a))=-\tan(\theta-\theta_a)\nonumber\\
&=-\frac{\tan\theta-\tan\theta_a}{1+\tan\theta\tan\theta_a}.
\end{align} 
\end{equation}  
Then, substituting (\ref{derive_RF2}) and (\ref{proof_h2}) into (\ref{proof_h4}), we obtain
\begin{equation}\label{proof_h5a}
\begin{align}
\tan\gamma_1&=\frac{\sqrt{c^2\tau^2-D^2\cos^2\theta}}{D|\cos\theta|}
\end{align} 
\end{equation}
or equivalently
\begin{equation}\label{proof_h5b}
\begin{align}
\gamma_1&=\arctan\left(\frac{\sqrt{c^2\tau^2-D^2\cos^2\theta}}{D|\cos\theta|}\right).
\end{align} 
\end{equation}
From (\ref{proof_h5b}) and by using some trigonometric identities, we can express the simplified $\sin\gamma_1$ and $\cos\gamma_1$ as, respectively,
\begin{equation}\label{proof_h6}
\begin{align}
\sin\gamma_1&=\sin\left(\arctan\left(\frac{\sqrt{c^2\tau^2-D^2\cos^2\theta}}{D|\cos\theta|}\right)\right)\nonumber\\
&=\frac{\sqrt{c^2\tau^2- D^2\cos^2\theta}}{c\tau},
\end{align} 
\end{equation}
and
\begin{equation}\label{proof_h7}
\begin{align}
\cos\gamma_1&=\cos\left(\arctan\left(\frac{\sqrt{c^2\tau^2-D^2\cos^2\theta}}{D|\cos\theta|}\right)\right)\nonumber\\
&=\frac{D|\cos\theta|}{c\tau}.
\end{align} 
\end{equation}
Finally, by substituting (\ref{proof_h6}) and (\ref{proof_h7}) into (\ref{proof_h1a}), we obtain 
\begin{equation}\label{proof_h8} 
\begin{align}
h_1=\frac{l\sqrt{c^2\tau^2- D^2\cos^2\theta}}{c\tau}+\frac{wD|\cos\theta|}{c\tau}.
\end{align} 
\end{equation}

Similarly, the region $UPQTZW$ is the combined area of two right-angled triangles $PQT$ and $UZW$, and a parallelogram $UPTZ$ with height denoted by $h_2$ and base $|R_1$Rx$|$, which is the length of the line segment $R_1$Rx. This area is given by 
\begin{equation}\label{eq:10b}
S_{UPQTZW}=\frac{lw}{2}+\frac{lw}{2}+|R_1\text{Rx}|\cdot h_2=lw+|R_1\text{Rx}|\cdot h_2
\end{equation}
where after the similar derivation as (\ref{proof_h8}), we have $h_2=h_1$.
Meanwhile, the region $PQTS$ is a rectangle, whose area can be calculated as
\begin{equation}\label{eq:10c}
S_{PQTS}=lw.
\end{equation}

Next, we calculate the area of the region $PSE$. Intuitively,  it can be seen that the area and shape of this region depend on the building orientation $\theta$. In one extreme case when $\theta=\pi/2$, this region is of a rectangle shape whose area is given by
\begin{equation}\label{eq:10d1}
S_{PSE}=\frac{l(c\tau-D)}{2}.\end{equation}
In the other extreme case when $\theta=\pi$, the shape of the region $PSE$ is a triangle, which is given by  
\begin{equation}\label{proof_triangle1}\begin{align}
S_{PSE}=\frac{tl}{2}
\end{align} \end{equation}
where $t$ is the height of the triangle. Note that $\angle{EPS}$ in Fig.~{\ref{fig:3}} is equal to $\gamma_1$. Thus we have 
\begin{equation}\label{proof_triangle2}\begin{align}
t&=\frac{l \tan\gamma_1}{2}.
\end{align}\end{equation}
From (\ref{proof_h5a}), for the case of $\theta=\pi$, we have
\begin{equation}\label{proof_triangle3}
\begin{align}
\tan\gamma_1&=\frac{\sqrt{c^2\tau^2-D^2\cos^2\theta}}{D|\cos\theta|}=\frac{\sqrt{c^2\tau^2-D^2}}{D}.
\end{align} 
\end{equation}
Substituting (\ref{proof_triangle2}) and (\ref{proof_triangle3}) into (\ref{proof_triangle1}), we have 
\begin{equation}\label{proof_triangle4}\begin{align}
S_{PSE}=\frac{l^2\sqrt{c^2\tau^2-D^2}}{4D}.
\end{align} \end{equation}

Meanwhile, for the general case of $\pi/2<\theta<\pi$, the region $PSE$ is a trapezium. Though we can analytically derive the exact area of such a region as a function of $\theta$, we find that the resultant expression is much complicated, which makes the subsequent derivation very complex. Thus, we present an approximate but sufficiently accurate expression for the area of the region $PSE$ by only taking average between the aforementioned two extreme cases, i.e.,
\begin{equation}\label{eq:10d}
S_{PSE}\approx\frac{l(c\tau-D)}{4}+\frac{l^2\sqrt{c^2\tau^2-D^2}}{8D}.\end{equation}

Finally, by substituting (\ref{eq:10a}), (\ref{eq:10b}), (\ref{eq:10c}) and (\ref{eq:10d}) into (\ref{eq:10}), we obtain the area of the blockage region $JKQTZWUEM$ as 
\begin{equation}\begin{flalign}\label{eq:11}
S_{JKQTZWUEM}&\approx(|\text{Tx}R_1|+|R_1\text{Rx}|)h_1+lw\nonumber\\
&~~-\frac{l(c\tau-D)}{4}-\frac{l^2\sqrt{c^2\tau^2-D^2}}{8D}\nonumber\\
&\stackrel{(a)}{=}c\tau h_1+lw-\frac{l(c\tau-D)}{4}-\frac{l^2\sqrt{c^2\tau^2-D^2}}{8D}
\end{flalign}\end{equation}
where the equality (a) holds because the total length of $|\text{Tx}R_1|$ and $|R_1\text{Rx}|$ is equal to the path length of the considered first-order reflection path $L_r=c\tau$. This completes the proof.\\
\subsection{Proof of \textbf{Theorem 2}}\label{proof_Theorem2}
Since the first-order reflection path will occur when there is no building whose center falls within the blockage region $JKQTZWUEM$, we have 
\begin{equation}\label{eq:16a}
f_{NB_1}(\tau|\theta)=\mathbb{P}(K_{NB_1}=0)=\frac{\left(\mathbb{E}[K_{NB_1}]\right)^0}{0!}e^{-\mathbb{E}[K_{NB_1}]}.\end{equation}
Substituting (\ref{eq:15}) into (\ref{eq:16a}), we can directly obtain the $f_{NB_1}(\tau|\theta)$ as in the Theorem 2.\\
\subsection{Derivation of (\ref{eq:20})}\label{proof_closedform}
We first substitute (\ref{eq:19a}) and (\ref{eq:19b}) into (\ref{eq:3}) and obtain
\begin{equation}\begin{align}\label{eq:19c}
f_{SR}(\tau)&=\frac{1}{\pi}\left(\int_0^{\pi/2}f_{SR}(\tau|\theta)\,d\theta+\int_{\pi/2}^{\pi} f_{SR}(\tau|\theta)\,d\theta\right).
\end{align}\end{equation}
We notice that when the value of $\varphi$ in the first term on the right hand side of (\ref{eq:19c}) is $\varphi=\pi-\theta$, the resulting $f_{SR}(\tau|\theta)$ and the $f_{SR}(\tau|\theta)$ in the second term on the right hand side of (\ref{eq:19c}) are symmetric about $\pi/2$. Thus, we can further simplify (\ref{eq:19c}) as
\begin{equation}\label{proof_closedform1}\begin{flalign}
&f_{SR}(\tau)=\frac{2}{\pi}\int_{\pi/2}^{\pi} f_{SR}(\tau|\theta)\,d\theta\nonumber\\
&\approx\frac{2}{\pi}\int_{\pi/2}^{\pi}\frac{\lambda \mathbb{E}[l] c\tau}{\sqrt{c^2\tau^2-D^2\cos^2\theta}} \cdot\nonumber\\
&\exp\biggl(-\lambda\Bigl(\mathbb{E}[l]\sqrt{c^2\tau^2-D^2\cos^2\theta}+\mathbb{E}[w]D\left|cos\theta\right|\nonumber\\
&+\mathbb{E}[l]\mathbb{E}[w]-\frac{\mathbb{E}[l](c\tau-D)}{4}-\frac{\mathbb{E}[l]^2\sqrt{c^2\tau^2-D^2}}{8D}\Bigr)\biggr) d\theta\nonumber\\
&+\frac{2}{\pi}\int_{\pi/2}^{\pi}\frac{\lambda \mathbb{E}[w] c\tau}{\sqrt{c^2\tau^2-D^2\cos^2\theta}} \cdot\nonumber\\
&\exp\biggl(-\lambda\Bigl(\mathbb{E}[w]\sqrt{c^2\tau^2-D^2\cos^2\theta}+\mathbb{E}[l]D\left|cos\theta\right|\nonumber\\
&+\mathbb{E}[w]\mathbb{E}[l]-\frac{\mathbb{E}[w](c\tau-D)}{4}-\frac{\mathbb{E}[w]^2\sqrt{c^2\tau^2-D^2}}{8D}\Bigr)\biggr)d\theta.\end{flalign}\end{equation}
Next, we let $a=c\tau/D$. For simplicity, we perform an approximation of $\left|\cos\theta\right|$ in the exponential function of (\ref{proof_closedform1}). The approximation expression can be written as
\begin{equation}\label{proof_closedform2} 
\begin{align}
\left|\cos\theta\right|&=\frac{(\sqrt{a^2-1}-a)\cos\theta + a}{\sqrt{a^2-1}-a}-\frac{a}{\sqrt{a^2-1}-a}\nonumber\\
&\approx\frac{\sqrt{a^2-\cos^2\theta}}{\sqrt{a^2-1}-a}-\frac{a}{\sqrt{a^2-1}-a}.
\end{align} 
\end{equation}
Therefore, by replacing (\ref{proof_closedform2}) into (\ref{proof_closedform1}), we have
\begin{equation}\label{proof_closedform3}\begin{align}
f_{SR}(\tau)\approx&\int_{\pi/2}^{\pi}\frac{\zeta_1'\exp(-b_1\sqrt{a^2-\cos^2\theta})}{\sqrt{a^2-\cos^2\theta}}\nonumber\\
&+\int_{\pi/2}^{\pi}\frac{\zeta_2'\exp(-b_2\sqrt{a^2-\cos^2\theta})}{\sqrt{a^2-\cos^2\theta}}
\end{align}\end{equation}
where
\begin{equation}\label{proof_closedform4}
\begin{align}
\zeta_1'&=\lambda\mathbb{E}[l]a\exp\biggl(-\lambda\mathbb{E}[l]\mathbb{E}[w]+\frac{\lambda\mathbb{E}[l]D(a-1)}{4}\biggr)\nonumber\\
&\times\exp\biggl(\frac{\lambda\mathbb{E}[l]^2\sqrt{a^2-1}}{8}+\frac{\lambda\mathbb{E}[w]Da}{\sqrt{a^2-1}-a}\biggr)\nonumber\\
b_1&=\lambda D \biggl(\mathbb{E}[l]+\frac{\mathbb{E}[w]}{\sqrt{a^2-1}-a}\biggr)\nonumber\\
\zeta_2'&=\lambda\mathbb{E}[w]a\exp\biggl(-\lambda\mathbb{E}[w]\mathbb{E}[l]+\frac{\lambda\mathbb{E}[w]D(a-1)}{4}\biggr)\nonumber\\
&\times\exp\biggl(+\frac{\lambda\mathbb{E}[w]^2\sqrt{a^2-1}}{8}+\frac{\lambda\mathbb{E}[l]Da}{\sqrt{a^2-1}-a}\biggr)\nonumber\\
b_2&=\lambda D \biggl(\mathbb{E}[w]+\frac{\mathbb{E}[l]}{\sqrt{a^2-1}-a}\biggr)\nonumber.
\end{align} 
\end{equation}
Next, we perform the first-order Taylor expansions for $\sqrt{a^2-\cos^2\theta}$, $\exp(-b_1\sqrt{a^2-\cos^2\theta})$ and $\exp(-b_2\sqrt{a^2-\cos^2\theta})$, separately  at $\theta=\frac{3\pi}{4}$. Their first-order Taylor expansions are given as
\begin{equation}\label{proof_closedform5}
\begin{align}
\sqrt{a^2-\cos^2\theta}&\approx\eta-\frac{3\pi}{8\eta}+\frac{\theta}{2\eta}
\end{align} 
\end{equation}
where $\eta=\sqrt{a^2-(1/2)}$,
\begin{equation}\label{proof_closedform6}
\begin{align}
\exp(-b_1\sqrt{a^2-\cos^2\theta})&\approx\exp(-b_1\eta)\left(1+\frac{3\pi b_1}{8\eta}\right)\nonumber\\
&-\frac{b_1\exp(-b_1\eta)}{2\eta}\theta,
\end{align} 
\end{equation}
and
\begin{equation}\label{proof_closedform7}
\begin{align}
\exp(-b_2\sqrt{a^2-\cos^2\theta})&\approx\exp(-b_2\eta)\left(1+\frac{3\pi b_2}{8\eta}\right)\nonumber\\
&-\frac{b_2\exp(-b_2\eta)}{2\eta}\theta.
\end{align} 
\end{equation}
Finally, substituting (\ref{proof_closedform5}), (\ref{proof_closedform6}) and (\ref{proof_closedform7}) into (\ref{proof_closedform3}), we obtain the closed-form approximation for the $f_{SR}(\tau)$ as in (\ref{eq:20}).

\ifCLASSOPTIONcaptionsoff
  \newpage
\fi

\bibliographystyle{IEEEtran}
\bibliography{TVT_2016}
\end{document}